\documentclass[a4paper,12pt]{jpconf}

\begin{document}

\title[Ageing without detailed balance]{Ageing without detailed balance:
local scale invariance applied to two exactly solvable models}

\author{Florian Baumann$^{a,b}$}
\address{$^a$Institut f\"ur Theoretische Physik I,
Universit\"at Erlangen-N\"urnberg, \\
Staudtstra{\ss}e 7B3, D -- 91058 Erlangen, Germany}
\address{$^b$Laboratoire de Physique des
Mat\'eriaux,\footnote{Laboratoire associ\'e au CNRS UMR 7556}
Universit\'e Henri Poincar\'e Nancy I, \\
B.P. 239, F -- 54506 Vand{\oe}uvre l\`es Nancy Cedex, France}

\begin{abstract}
I consider ageing behaviour in two exactly solvable
reaction-diffusion systems. Ageing exponents and scaling
functions are determined. I discuss in particular a case in which
the equality of two critical exponents,
known from systems with detailed balance, 
does not hold any more. Secondly it is shown that the form of the
scaling functions can be understood by symmetry considerations.
\end{abstract}


\section{Introduction}

Ageing phenomena may occur in systems which are rapidly
quenched into a region in parameter space with several
competing stationary states. These phenomena have been studied
quite extensively in systems with detailed balance such 
as simple magnetic systems, which are initially prepared in a
high-temperature state and then quenched to or below the critical
temperature $T_C$
\cite{Bray94,Cate00,Cugl02,Godr02,Henk04,Calabrese}. 
One typically considers the autocorrelation and the
autoresponse function, for which one expects scaling behaviour
in the {\sl ageing regime}, that is for $t$,$s$ and $t-s$ large 
compared to some microscopic timescale:
\begin{equation}
\label{definitions}
\begin{array}{ccccc}
   C(t,s) & :=& \langle \phi(\mathbf{x},t) \phi(\mathbf{x},s) \rangle & \sim & s^{-b} f_C
   (t/s) \\
   R(t,s) & :=& \frac{\delta \langle \phi{(\mathbf{x},t}) \rangle} {\delta
   h(\mathbf{x},s)} & \sim & s^{-a-1} f_R(t/s)
\end{array}
\end{equation}
where $\phi(\mathbf{x},t)$ is the order parameter describing the
system and $h(\mathbf{x},s)$ is a small
external perturbation, for instance a magnetic field. 
$a$ and $b$ are critical exponents and the scaling functions
$f_C$ and $f_R$ behave for large arguments as
\begin{equation}
   \label{scaling}
   \begin{array}{ccc}
     f_C(y) & \stackrel{y \rightarrow \infty}{\sim} & y^ {-\lambda_C/z} \\
     f_R(y) & \stackrel{y \rightarrow \infty}{\sim} & y^ {-\lambda_R/z}
   \end{array}
\end{equation}
where $\lambda_R$ and $\lambda_C$ are new exponents and $z$ is 
the dynamical critical
exponent. For systems with decorrelated initial conditions $\lambda_R =
\lambda_C$ has been found and the relation $a = b$ has been confirmed
at criticality.
Notice that for systems with detailed balance the latter condition
is necessary because the fluctuation-dissipation ratio has to
hold when the system reaches equilibrium.

The approach of local scale invariance (LSI) had been proposed in
\cite{Henk02,Pico04} to understand the form of $C(t,s)$ and
$R(t,s)$ on the basis of 
symmetry considerations. It
turned out that for magnetic systems 
the form of $R(t,s)$ can be completely fixed
by symmetries whereas $C(t,s)$ is fixed up to a scaling
function. Monte Carlo simulations in the
Ising model and the XY model showed excellent
agreement with these predictions
\cite{henkel1,henkel2,henkel3,abriet}, 
whereas renormalisation group computations yielded a
correction at
two loops \cite{calabrese2}, the origin of which is still not quite clear.  

In the systems considered so far detailed balance holds. However
many realistic systems do not possess this property. It
is therefore interesting to see what happens if this
condition is relaxed. Typical systems which lack detailed 
balance are reaction-diffusion
systems, where particles undergo diffusion on a lattice and in
addition, there are particle
creation and annihilation processes. Numerical work has been done  
on the (fermionic)
contact process \cite{Enss04,Rama04}, showing that at criticality
dynamical scaling holds for the response function and the {\sl
connected} correlator, but that opposed to the above mentioned
magnetic systems $a \neq b$. Here, I shall ask the question: 
Are there exactly solvable systems without detailed balance 
where $a \neq b$, and if so, can
the form of the scaling functions still be understood with 
the help of the theory of local scale invariance?  

In this text I shall consider two specific models without
detailed balance. On the one hand they allow for exact calculation
of the ageing exponents and the scaling functions, on the
other hand they can be described by a field-theoretical formalism
and can therefore be attacked by the theory of local scale
invariance. This paper is organised as follows: In section
2, I introduce the models and present the main results. In
particular I show that in one case one indeed encounters $a \neq b$. 
In section 3, I look at the same
models from the field-theoretical perspective and
demonstrate that the
form of the scaling functions can indeed be understood by
considering the symmetries of the models.

\section{The bosonic contact and pair-contact processes}
\vspace{0.5cm}
\subsection{The models}

I consider a $d$-dimensional cubic lattice with one sort of particles $A$. 
It is important that the system is {\sl bosonic}, which means that 
there is {\sl no restriction} on the number of particles on one lattice site. 
In what follows I shall consider two different models:
\begin{itemize}
   \item \underline{The bosonic contact process (BCPD):} The particles undergo
   diffusion with diffusion constant $D$, that is, they may
   jump to a nearest-neighbor site with rate $D$. Furthermore 
   single particles can disintegrate with
   rate $\lambda$ or produce offspring with rate $\mu$. The
   newly produced particles then sit on the same lattice
   site as the original particle. 
   \[
   A \stackrel{D}{\longleftrightarrow} A, \quad  A
   \stackrel{\lambda}{\longrightarrow} \emptyset, \quad A
   \stackrel{\mu}{\longrightarrow} 2 A
   \]
   This process has been used to model the clustering of biological organisms
   in \cite{Houc02}.
   \item \underline{The bosonic pair-contact process (BPCPD):} Also in this
   process there is diffusion with constant $D$. In addition
   two particles on one lattice site can
   coagulate with rate $\lambda$ or give birth to a third particle with rate
   $\mu$:
   \[
   A \stackrel{D}{\longleftrightarrow} A, \quad 2 A
   \stackrel{\lambda}{\longrightarrow} A, \quad 2 A
   \stackrel{\mu}{\longrightarrow} 3 A
   \]
   It has been shown by Paessens and Sch{\"u}tz
   \cite{Paes04a} that this model
   can be solved analytically for any finite time.
   This is mainly due to a formal analogy to the
   spherical model, where the {\sl control parameter}
   $\alpha$ defined below formally replaces the temperature.
\end{itemize}

The state of the system is characterised by the number of particles on each
site. I denote this by  $\{n\} =
\{\ldots,n_{\mathbf{x}},\ldots \}$,
where the non-negative integer $n_{\mathbf{x}}$ gives the number
of particles on site $\mathbf{x}$. The temporal evolution can be 
described by a master equation, which can be turned into a Schr{\"o}dinger-type 
equation by standard techniques
\cite{Doi76,Peliti,Taeub05}. One then has annihilation and creation 
operators $a(\mathbf{x})$ and $a^\dag (\mathbf{x})$ at each
lattice site, the nonvanishing commutation relations of
which are $[a(\mathbf{x}),a^\dag(\mathbf{y})] =
\delta_{\mathbf{x},\mathbf{y}}$. Then two quantities are defined: the state vector $|n\rangle
:= \prod_{\mathbf{x}} (a^\dag(\mathbf{x}))^{n_{\mathbf{x}}}
|0\rangle$, where $|0\rangle$ is the vacuum state
representing the empty lattice, and the
vector $|P(t)\rangle := \sum_{\{n\}}
P(\{n\},t) | n \rangle$, where $P(\{n\},t)$ is the
probability to find the system in the state $\{n\}$ at time
$t$. The latter quantity obeys the equation $ \partial_t |P(t)
\rangle = - H |P(t) \rangle$, where
the Hamiltonian $H$ is given in terms of annihilation and
creation operators and can be found in \cite{Paes04a} for the
cases at hand. Time-dependent observables are obtained by 
passing to the Heisenberg picture and the temporal evolution 
of an observable $g(t)$ is then given by the Heisenberg equation of motion
\begin{equation}
   \label{eqn:heisenberg}
   \partial_t g(t) = [H,g(t)]
\end{equation}
Finally, the average of $g(t)$ is
calculated as $\langle g \rangle (t) := \langle s | g(t) | P(0) \rangle$, where
$\langle s|$ is a coherent state vector with the properties
$\langle s | a^\dag(\mathbf{x}) = \langle s |$ and $
\langle s | H = 0$. The quantities of interest are:
\begin{itemize}
   \item \underline{The local particle density}
     \begin{equation}
        \label{part_den}
        \rho(\mathbf{x},t) := \langle a^\dag(\mathbf{x},t)
	a(\mathbf{x},t) \rangle = \langle a(\mathbf{x},t)
	\rangle 
     \end{equation} 
   where the special property of the state $\langle s |$ has
   been used in the last equality.
   \item \underline{The connected two-point correlator:}
     \begin{equation}
        \label{two_point}
        G(\mathbf{x}-\mathbf{y},t,s) := \langle
	a(\mathbf{x},t) a(\mathbf{y},s) \rangle - \langle
	a(\mathbf{x},t)\rangle \langle a(\mathbf{x},s)
	\rangle
     \end{equation}
   Here and in what follows I assume spatial translation
   invariance, so that two-point quantities depend only on
   the difference of the spatial coordinates. If scaling
   behaviour is found, the corresponding scaling
   function will be denoted by $f_G$ in analogy to (\ref{definitions}).
   \item \underline{The response function:} To compute this
   quantity, one adds a small
   perturbation $\sum_{\mathbf{x}} h(\mathbf{x},t)
   a^\dag(\mathbf{x})$ to the Hamiltonian $H$. This
   corresponds to spontaneous particle creation at an empty
   lattice site with rate $h(\mathbf{x},t)$. Then the
   response function is simply defined as
   \begin{equation}
      \label{response}
      R(\mathbf{x}-\mathbf{y},t,s) := \frac{\delta \langle
      a(\mathbf{x},t) \rangle }{\delta h(\mathbf{y},s) }
   \end{equation}
\end{itemize}
The Heisenberg equation of motion (\ref{eqn:heisenberg}) is
used to derive differential equations for the quantities
(\ref{part_den})-(\ref{response}). For the particle density and
the two-point correlator the following initial condition are
chosen \footnote{This ensures that the system is
translationally
invariant. Furthermore it can be shown that this choice
can be realised with an initial Poisson distribution of the particle
density at each lattice site.}
\begin{equation}
 \label{eqn:intitial_cond}
 \rho(\mathbf{x},0) = \rho_0, \quad G(\mathbf{x},0,0) = 0
\end{equation}
whereas the response function is required to be a delta-peak at
$t = s$. All these equations can be solved by standard
techniques.

\subsection{Results}

For the particle density, one finds the following results for
both processes \cite{Houc02,Paes04a}
\begin{itemize}
   \item For $\mu > \lambda$ particle creation outweighs
   particle annihilation and $\rho(\mathbf{x},t)$ diverges.
   \item For $\mu < \lambda$ particle annihilation is
   stronger and the systems runs into the empty lattice
   state.
   \item Only if $\mu = \lambda$ creation and annihilation
   of particles are of equal strength. In this case one has
   $\rho(\mathbf{x},t) = \rho_0$ for all times.
\end{itemize}

In the sequel I shall
only look at the most interesting case when $\lambda =\mu$.
Then the creation and annihilation processes balance
each other, one says that one is on the {\sl critical line}.
As the particle density remains constant, 
one needs to look at the variance $\sigma_\rho(t)$ of
$\rho(\mathbf{x},t)$ for
further insights. Notice that $\sigma_\rho(t)$ is equal to
$G(\mathbf{0},t,t)$ up to a constant. One finds the
following results \cite{Houc02,Paes04a}
\begin{itemize}
  \item For the BCPD the correlator $G(\mathbf{0},t,t)$ behaves as 
  $t^{-\frac{d}{2}+1}$ (for $t \rightarrow \infty$) and one
  has a diverging variance
  $\sigma_\rho(t)$ if $d < 2$ \footnote{In the case $d = 2$
  there is a logarithmic divergence.}, otherwise the variance
  is bounded. This means that for $d \leq 2$ diffusion can
  not spread particles evenly so that they accumulate on
  very few lattice sites - a {\sl clustering
  transition} occurs
  \item For the BPCPD the control parameter $\alpha$ is
  defined by
  \begin{equation}
   \alpha := \mu/D.
  \end{equation}
  It measures the strength of the creation and annihilation
  processes in comparison to the diffusion process. It turns out that
  there is a critical value $\alpha_C > 0$ and the behaviour
  of $G(\mathbf{0},t,t)$ depends on whether $\alpha$ is larger, equal
  or smaller than $\alpha_C$. More precisely
  \begin{equation}
  \label{cases}
   \begin{array}{c|c|c|c}
     \alpha < \alpha_C, d>2 & \alpha = \alpha_C, d > 4 &
      \alpha = \alpha_C, 2 < d < 4 & \alpha > \alpha_C \;
          \mbox{or} \; d \leq 2 \\ \hline
     G(\mathbf{0},t,t) \stackrel{t \rightarrow
     \infty}{\longrightarrow} const &
     G(\mathbf{0},t,t) \stackrel{t \rightarrow \infty}{\sim} 
     t & G(\mathbf{0},t,t) \stackrel{t \rightarrow \infty}{\sim}
     t^{\frac{d}{2}-1} & G(\mathbf{0},t,t) \stackrel{t \rightarrow
     \infty}{\sim} \exp(t/\tau) \\
   \end{array}
  \end{equation}
  For small $\alpha$, diffusion is dominant and the
  system stays more or less homogeneous. But if $\alpha$ is
  large enough, there is again a clustering transition.
\end{itemize}

Finally the two-time quantities \cite{baum} are considered. For the
response function scaling behaviour is found in the ageing
regime with the result
\begin{equation}
\label{result_r}
R(t,s) = r_0 s^{-\frac{d}{2}+1} \left( (t/s)^{-\frac{d}{2}+1} - 1\right)
\end{equation}
from which the critical exponents $a$ and $\lambda_R$  can be derived,
as well as the scaling function $f_R(y)$

For the connected two-time correlator, one also finds 
the scaling behaviour (\ref{scaling}) in the BCPD, and in the 
BPCPD for the first three cases discussed in (\ref{cases}). The ageing
exponents can be found in table (\ref{tb:results_exp}),
whereas the scaling function is given by the integral expression
\begin{equation}
\label{result_sf}
f_G(y) = g_0 \int_0^1 d \theta \; \theta^{(a-b)} (y + 1 - 2
\theta)^{-\frac{d}{2}} 
\end{equation}
Note that $a \neq b$ is found at criticality
\cite{baum}. This
entails in particular that there is no non-trivial analogue
to the fluctuation-dissipation ratio known from magnetic
system and that it is not possible to define in a
straightforward way an
effective temperature characterizing the system  as
suggested in
\cite{Sast03}.
 
\section{Local scale invariance}

\vspace{0.4cm}

In this last section I consider the same processes from a
field-theoretical point of view. First I define the fields
$\phi(\mathbf{x},t) := a(\mathbf{x},t) - \rho_0$ and
$\tilde{\phi} := a^\dag(\mathbf{x},t) - 1$. In this way the
correlator $\langle \phi(\mathbf{x},t) \phi(\mathbf{x}',s)
\rangle$ equals the connected correlator (\ref{two_point}).
Then, by taking the continuum limit, 
the Hamiltonian $H$ is turned into a field-theoretical 
action $\Sigma[\phi,\tilde{\phi}]$ \cite{Taeub05,Howa97},
from which $n$-point correlators can be computed via
\begin{equation}
 \langle \phi_1(\mathbf{x}_1,t_1) \ldots \phi_n(\mathbf{x}_n,t_n) \rangle =
 \int \mathcal{D}[\phi] \mathcal{D}[\tilde{\phi}]
 \phi_1(\mathbf{x}_1,t_1) \ldots \phi_n(\mathbf{x}_n,t_n)
 \exp(-\Sigma[\phi,\tilde{\phi}]).
\end{equation}
The action can be split up into two parts as $\Sigma[\phi,\tilde{\phi}] 
= \Sigma_0[\phi,\tilde{\phi}] +
\Sigma_{noise}[\phi,\tilde{\phi}]$,
where the first part $\Sigma_0[\phi,\tilde{\phi}]$ is given by 
\begin{equation}
 \begin{array}{ccll}
 \Sigma_0[\phi,\tilde{\phi}] & = & \int d \,\mathbf{R} \, d
 \,u \; [ \tilde{\phi} (2 \mathcal{M} 
 \partial_u - \nabla^2) \phi] & \qquad  \mbox{for the BCPD} \\
 \Sigma_0[\phi,\tilde{\phi}] & = & \int d \,\mathbf{R}\, d
 \, u \; [ \tilde{\phi} (2 \mathcal{M} \partial_u - \nabla^2) \phi - \alpha
 \tilde{\phi}^2 \phi^2] &\qquad \mbox{for the BPCPD}.
 \end{array}
\end{equation}
\begin{table}[h]
  \[
  \begin{array}{||c|c||c|c||}  \hline \hline
    & & \multicolumn{2}{c||}{\mbox{bosonic pair-contact process}} \\
         \cline{3-4}
    & \raisebox{1.5ex}[-1.5ex]{ \mbox{bosonic contact process}}
    & \alpha < \alpha_C  & \alpha = \alpha_C  \\ \hline
    \hline
    a  & \frac{d}{2}-1 & \frac{d}{2}-1 & \frac{d}{2}-1  \\
    \hline
    b & \frac{d}{2}-1& \frac{d}{2}-1  & \begin{array}{ccl}
      0 & \mbox{if} & 2 < d < 4  \\
      \frac{d}{2}-2 & \mbox{if} & d > 4
    \end{array}
      \\
    \hline \hline
  \end{array}
  \]
\caption[AgeingTab1]{Ageing exponents of the critical bosonic contact
and pair-contact processes in the different regimes. $\lambda_R = \lambda_G = d$ and $z
= 2$ was found for all cases.
The results for the bosonic contact process hold for an arbitrary dimension $d$, but for the
bosonic pair-contact process they only apply if $d>2$, since $\alpha_C=0$
for $d\leq 2$.}
\label{tb:results_exp}
\end{table}
Both fields under the integral depend on the
integration variables $\mathbf{R}$ and $u$, but here and in
what follows I shall suppress the arguments of the fields 
for space considerations.
$\mathcal{M}$ is a parameter related to the diffusion
constant. The exact form of the second part is somewhat more involved and can be
found in \cite{Howa97,baum2}. Here it suffices to state
that $\Sigma_{noise}[\phi,\tilde{\phi}]$ has the form
$\Sigma_{noise}[\phi,\tilde{\phi}] = \sum_{n,m \leq 3} 
a_{nm} \int d \mathbf{R} d u
\phi^n \tilde{\phi}^m$, where the constants
$a_{nm}$ vanish whenever $n \geq m$.
The point of this split-up is
the following: It can be shown that
in {\sl both cases}, $\Sigma_0[\phi,\tilde{\phi}]$ has
nontrivial symmetry properties. This is a well-known fact
for the BCPD \cite{Pico04} as the corresponding evolution equation for
$\phi$ is a free Schr{\"o}dinger equation, but has only been
shown recently \cite{stoimenov,baum2} for the case of the
BPCPD. From these symmetry properties the so-called Bargmann
superselection rule can be inferred, stating that $ \langle
\underbrace{\phi \ldots \phi}_n \underbrace{\tilde{\phi} \ldots
 \tilde{\phi}}_m \rangle_0 = 0$ unless $n = m$\footnote{For the BCPD and an even number of fields,
 this can also be inferred from causality combined with Wick's theorem}.
 \vspace{0.4cm}
This entails that the response function is
independent of $\Sigma_{noise}[\phi,\tilde{\phi}]$
\cite{Pico04,baum2} and is given by
\begin{equation}
  \label{response_twopoint}
  R(\mathbf{x}-\mathbf{x}',t,s) = \langle \phi(\mathbf{x},t)
  \tilde{\phi}(\mathbf{x}',s) \rangle_0
\end{equation}
where $ \langle \ldots \rangle_0$ denotes the average with
respect to $\Sigma_0[\phi,\tilde{\phi}]$. Further
investigations reveal, that the connected
correlator is essentially given by a sum of
integrals over three- and four-point functions, which have
also been computed with respect to
$\Sigma_0[\phi,\tilde{\phi}]$. Therefore, the main task is
to determine the two-, three-, and four-point functions of
the theory described by $\Sigma_0[\phi,\tilde{\phi}]$ only.

Let us consider the BCPD first. In this case the evolution equation
for the field $\phi$ is a free Schr{\"o}dinger equation.
The symmetry group of this equation, i.e. the group of
transformations carrying solutions to other solutions,
is the well-known Schr{\"o}dinger group \cite{Niederer}.
An element $g$ of this group acts on space-time coordinates as
$(\mathbf{x},t) \rightarrow (\mathbf{x}',t') = g(\mathbf{x},t)$
with
\begin{equation}
t \rightarrow t' = \frac{\tilde{\alpha} t + \beta}{\gamma t + \delta},
\;\;\;
\mathbf{x} \rightarrow \mathbf{x}' =
\frac{\mathcal{R} \mathbf{x} + \mathbf{v} t +
\tilde{\alpha}}{\gamma t + \delta}; \;\;\; 
\tilde{\alpha} \delta - \beta \gamma =
1
\end{equation}
where $\mathcal{R}$ ist a rotation matrix and
$\tilde{\alpha},\beta,\gamma,\delta$ and $\mathbf{v}$ are
parameters. A solution $\Psi(\mathbf{x},t)$ of the free
Schr{\"o}dinger equation is transformed as
\begin{equation}
\hspace{-1.5cm}
\Psi(\mathbf{x},t) \rightarrow (T_g\Psi)(\mathbf{x},t) =
f_g[g^{-1}(\mathbf{x},t)] \Psi[g^{-1}(\mathbf{x},t)] \quad
\end{equation}
where the companion function $f_g$ is known explicitely. A
field transforming in this way is called {\sl quasiprimary} and the important
assumption is to identify the fields $\phi$ and
$\tilde{\phi}$ as the appropriate quasiprimary fields of the theory. 
One can show that
$n$-point functions build from quasiprimary fields satisfy
certain linear
partial differential equations involving the generators of the
Schr{\"o}dinger group.
Solving these equations, one finds that the response
function can be fixed completely with the result
\cite{baum2,henkel}
\begin{equation}
R(\mathbf{x} - \mathbf{x}',t,s) = r_0 (t-s)^{\frac{1}{2}(x_1 +
x_2)} \left( \frac{t}{s} \right)^{\frac{1}{2}(x_1 - x_2)}
\exp \left(- \frac{\mathcal{M}}{2}
\frac{(\mathbf{x}-\mathbf{x}')^2}{t-s} \right)
\end{equation}
where $x_1$ and $x_2$ are free parameters which can
be adjusted so that this expression is in line with the result
(\ref{result_r}). To deduce the exact functional form of
$R(\mathbf{x}-\mathbf{x}',t,s)$ is the actual predictive power
of this theory, providing new information which can for
instance not be obtained by simple scaling arguments. 
Also three- and four-point functions can be fixed to a certain 
degree by the symmetries.  There is, however, a degree of
freedom that remains in the form of undetermined
scaling functions. One can show that the result (\ref{result_sf})
can be reproduced by a suitable choice of these functions
\cite{baum2}, but I shall not give the results here for
limitations of space.\\

The procedure works similarly for the BPCPD. However, here the
Schr{\"o}dinger equation one has to consider is non-linear,  
which leads to a modification of the generators of the symmetry group, see
\cite{stoimenov,stoimenov2}.  
The result for the response function in this case is \cite{baum2}
\begin{equation}
R(\mathbf{x} - \mathbf{x}',t,s) = r_0 (t-s)^{\frac{1}{2}(x_1 +
x_2)} \left( \frac{t}{s} \right)^{\frac{1}{2}(x_1 - x_2)}
\exp \left(- \frac{\mathcal{M}}{2}
\frac{(\mathbf{x}-\mathbf{x}')^2}{t-s} \right) \Psi
\left(\frac{t}{s} \cdot
\frac{t-s}{\alpha^{1/\hat{y}}}, \frac{\alpha}{(t-s)^{\hat{y}}} \right)
\end{equation}
with an arbitrary function $\Psi$ and another parameter
$\hat{y}$. Also this expression can be
brought into agreement with (\ref{result_r}). One can see
that for the BCPD the symmetries fix completely the form of the
response function whereas for the BPCPD an arbitrary scaling
function remains. We can argue however, that the universal
scaling function $\Psi$ should not depend on the
non-universal quantity $\alpha$, which leads to the
statement $\Psi(\frac{t}{s} \cdot
\frac{t-s}{\alpha^{1/\hat{y}}}, \frac{\alpha}{(t-s)^{\hat{y}}}) 
= \tilde{\Psi}(t/s)$ for another
undetermined function $\tilde{\Psi}$. This is in line with
predictions obtained by renormalisation group arguments
\cite{Calabrese}. Finally one can show that in the BPCPD also the result
(\ref{result_sf}) can be reproduced correctly by appropriately
adjusting the free parameters of the theory \cite{baum2}.
Again I can not treat the latter point more explicitly here
but have to defer the reader to the quoted references.\\

I conclude by summing up the main results of this paper. 
I have considered the bosonic 
contact and pair contact processes
on the critical line. These are typical processes without
detailed balance the absence of which leads to the violation of 
the relation $a = b$. However, also
for these systems without detailed balance the approach of local scale
invariance is still suitable to gain insight in the form of the
scaling functions, even though the symmetries do not suffice,
especially in the case of the pair-contact process, to fix
completely the quantities of interest.

\ack
I acknowledge the support by the Deutsche
Forschungsgemeinschaft through grant no. PL 323/2.
This work was also supported by the Bayerisch-Franz\"osisches
Hochschulzentrum (BFHZ). Furthermore I thank M. Henkel and
M. Pleimling for useful comments. \vspace{0.3cm}
\setcounter{footnote}{0}

\end{document}